\documentclass[preprint,12pt,authoryear]{elsarticle}

\usepackage{amssymb}

\usepackage{colortbl}
\usepackage{amsmath}
\usepackage{amsthm}
\usepackage{amsfonts}
\usepackage{graphicx}
\usepackage{subfigure}

\usepackage{booktabs}
\usepackage{hyperref}

\usepackage{newfloat}
\usepackage{listings}
\usepackage{newfloat}
\usepackage{caption,setspace}

\journal{Neurocomputing}

\begin{document}


\begin{frontmatter}


\newcommand{\eat}[1]{}
\newcommand{\hao}[1]{{\color{blue}{#1}}}
\newcommand{\can}[1]{{\color{red}{#1}}}
\newcommand{\TODO}[1]{{\color{red}TODO:{#1}}}
\newcommand\beftext[1]{{\color[rgb]{0.5,0.5,0.5}{BEFORE:#1}}}
\newtheorem{pro}{Problem}
\newtheorem{defi}{Definition}
\title{Dual-space Hierarchical Learning for Goal-guided Conversational Recommendation}

\author[Can Chen]{Can Chen}

\affiliation[Can Chen]{
            organization={McGill University, Mila - Quebec AI Institute},
            country={Canada}}




\author[Hao Liu]{Hao Liu\texorpdfstring{\corref{cor1}}{}}
\ead{liuh@ust.hk}
\affiliation[Hao Liu]{
    organization={The Hong Kong University of Science and Technology (Guangzhou)},
    country={China}}
\cortext[cor1]{Corresponding author}

\author[Zeming Liu]{Zeming Liu}

\affiliation[Zeming Liu]{organization={Harbin Institute of Technology}, country={China}}

\author[Xue Liu]{Xue Liu}

\affiliation[Xue Liu]{organization={McGill University},country={Canada}}

\author[Dejing Dou]{Dejing Dou}
\affiliation[Dejing Dou]{organization={Baidu Research}, country={China}}



\begin{abstract}
Proactively and naturally guiding the dialog from the non-recommendation context~(e.g., Chit-chat) to the recommendation scenario \textcolor{black}{(e.g., Music)} is crucial for the Conversational Recommender System~(CRS).
Prior studies mainly focus on planning the next dialog goal~(e.g., chat on a movie star) conditioned on the previous dialog.
However, we find the dialog goals can be simultaneously observed at different levels, which can be utilized to improve CRS.
In this paper, we propose
\textit{\textbf{D}ual-space \textbf{H}ierarchical \textbf{L}earning}~(\textbf{DHL})
to leverage multi-level goal sequences and their hierarchical relationships for conversational recommendation.
Specifically, we exploit multi-level goal sequences from both the representation space and the optimization space.
In the representation space, we propose the hierarchical representation learning where a cross attention module derives mutually enhanced multi-level goal representations.
In the optimization space, we devise the hierarchical weight learning to reweight lower-level goal sequences, and introduce bi-level optimization for stable update.
Additionally, we propose a soft labeling strategy to guide optimization gradually.
Experiments on two real-world datasets verify the effectiveness of our approach.
Code and data are available \href{https://github.com/GGchen1997/NeuroComputing}{here}. 
\end{abstract}


\end{frontmatter}






\section{Introduction}

Recent years have witnessed the fast development of the Conversational Recommender System~(CRS)~\cite{sun2018conversational, christakopoulou2018q, zhang2018towards, yu2019visual, lei2020estimation, lei2020interactive}, which aims to recommend proper items through human-machine natural language interactions.
Compared with traditional recommender systems which rely on historical logs, CRS captures dynamic user interests by interacting with users in a more free-form way
(i.e. asking questions or recommending items).
Therefore, CRS has been widely adopted for various recommendation scenarios, including e-commerce, search engine, and virtual assistant.
\begin{figure}[t]
\centering
    \includegraphics[width=.7\columnwidth]{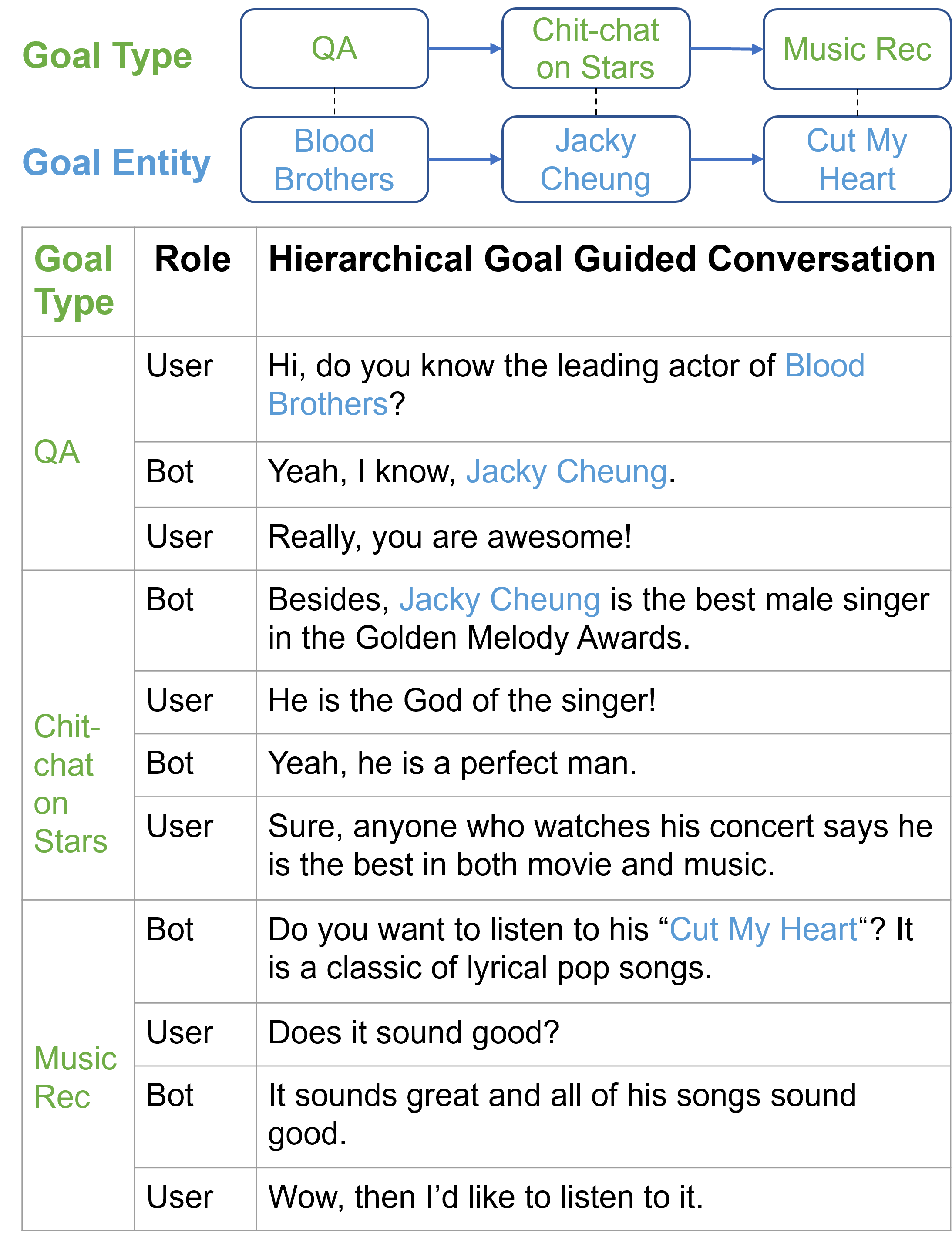} 
    \caption{
    An example of muti-level goal sequences in a human-machine conversation.
    }
    \label{fig: intro}
\end{figure}

Rather than assuming users always bear in mind what they want, one emerging direction~\cite{liu-etal-2020-towards-conversational, zhou2020topicguided} in CRS is to explore proactively discovering users' interests and naturally leading the conversation from the non-recommendation context to the recommendation scenario.
As illustrated in Figure~\ref{fig: intro}, the CRS can intelligently lead the goal type from \textit{QA} to \textit{Chit-chat on Stars} and finally reach the goal of \textit{Music Rec}.
To achieve this transition, the study in \cite{liu-etal-2020-towards-conversational} explicitly constructs dialog goal sequences and proposes a CNN based model to plan the next goal.
Besides, the study in \cite{zhou2020topicguided} combines the sequential recommendation model and the pre-trained language model to guide topic transitions by leveraging various conversation signals.
In summary, such methods learn a single representation of previous goals which could then guide the conversation towards the final recommendation goal.

However, after exploring large-scale recommendation dialogs, we observe the sequence of dialog goals co-exists in multiple levels.
Take the dialog in Figure~\ref{fig: intro} again for example, we can observe the sequence of dialog goals in at least two levels: the goal type level and the goal entity level.
To be specific, the goal type of the dialog first transits from \textit{QA} to \textit{Chit-chat on Stars} and then reaches \textit{Music Rec}. 
The corresponding goal entity sequence can be summarized as \textit{Blood Brothers} $\rightarrow$ \textit{Jacky Cheung} $\rightarrow$ \textit{Cut My Heart}.
Such multi-level dialog goal sequences are correlated and contain an intrinsic dependency hierarchy, which can be leveraged to enhance the next goal planning capability and improve the
effectiveness of CRS.

To this end, we propose the \textit{\textbf{D}ual-space \textbf{H}ierarchical \textbf{L}earning}~(\textbf{DHL}) to exploit multi-level goal sequences for proactive and natural conversational recommendation.
Specifically, DHL models the hierarchical and sequential structure of dialog goals from two spaces.
In the representation space, we propose the hierarchical representation learning where a cross attention module captures the hierarchical dependency between multi-level goal sequences.
In particular, the cross attention module consists of two symmetric components, \textit{type2entity} and \textit{entity2type}, to derive mutually enhanced representations of multi-level goal sequences.
In the optimization space, a hierarchical weight learning module is introduced to reweight goal sequences based on intermediate prediction results for better information use.
More specifically, higher accuracy of high-level prediction indicates more useful information for low-level goal planning task, which is assigned with a larger hierarchical weight.
To avoid trivial solutions of joint model parameter and hierarchical weight optimization, we introduce bi-level optimization~\cite{franceschi2018bilevel} to distill weak supervision signals from the training data for the stable update of hierarchical weights.
Additionally, we propose a soft labeling strategy to guide the dialog to the final recommendation goal gradually.
By assigning a small parameter of the final recommendation goal to each one-hot encoded current goal, the soft label can incorporate the global optimization direction information in the model training phase.
In summary, we make the following three major contributions.
\begin{itemize}
    \item To model hierarchical relationships between multi-level goal sequences in the representation space, we propose the hierarchical representation learning in which a cross attention module derives multi-grained goal representations in a mutual reinforcement way. 
    \item To leverage the hierarchical structure of goal sequences in the optimization space, we develop the hierarchical weight learning to adaptively reweight multi-level goal planning tasks. The bi-level optimization is introduced to stabilize the update of hierarchical weights.
    \item To guide conversation to the final recommendation goal, we propose a novel soft labeling strategy to adjust the global optimization direction.
\end{itemize}
We have conducted extensive experiments on two real-world conversational recommendation datasets, and the results demonstrate the effectiveness of our proposed approach.

\section{Preliminaries}

We first introduce some important definitions and then formalize the problem we aim to investigate.

Let $D=\{d_k\}_{k=1}^{N_d}$ denote a set of dialogs, where $N_d$ is the total number of dialogs in the dataset.
Each dialog $d_k\in D$ consists of multiple utterances between user and machine.

\begin{defi}{\textbf{Goal}.}
\label{def:goal} 
A goal $g$ is defined as the topic or knowledge~(e.g., an event, a movie star, etc.) the utterances focus on to keep the conversation natural and engaging.
\end{defi}

Depend on the granularity, the goal can be defined at different levels. 
In this paper, we define dialog goals in three levels, (1) goal type $g_p$ describes the type of a sub-dialog (e.g., QA, Chit-chat and recommendation), (2) goal entity $g_e$ means some entity the sub-dialog focuses about (e.g., some movie star and music), and (3) goal attribute $g_r$ describes goal entity~\textcolor{black}{(e.g., descriptions about the star like "the best male singer").}
Some consecutive utterances may share the same goal.
We use $N_p$, $N_e$ and $N_r$ to denote the number of goal type, goal entity, and goal attribute in $D$.

\begin{defi}{\textbf{Goal sequence}.}
\label{def:goalseq}
Given a dialog $d_i$, the goal sequence $\mathbf{g}_i=[g_1, g_2, \dots, g_f]$ is defined as a knowledge path that describes the semantic transition of topics in $d_i$.
\end{defi}

{Note that $g_f$ denotes the final recommendation goal, which is predefined before the dialog. 
The final recommendation goal setting~\cite{liu-etal-2020-towards-conversational, zhou2020topicguided} has wide applications since the real-world recommendation aims to deliver user interesting products/contents, which is usually predefined in customer/audience targeting and advertising.}
To ensure consistency, we constrain all goals in a goal sequence at the same level.
We denote $\mathbf{g}^p$ as the goal type sequence, $\mathbf{g}^e$ as the goal entity sequence, and $\mathbf{g}^r$ as the goal attribute sequence.
Consider a goal sequence $[g_1, g_2, \dots, g_t]$ in a particular level, denote the fixed-length initialization for each goal as $[\mathbf{x}_1, \mathbf{x}_2, \dots, \mathbf{x}_t]$.

\begin{defi}{\textbf{Adjacency matrix}.}
The adjacency matrix $\mathbf{C}^{pe}\in \mathbb{R}^{N_p \times N_e}$
captures the co-occurrence relationship between the goal type and the goal entity, where $c^{pe}_{ij} \in C^{pe}$ between goal type $g^p_i$ and goal entity $g^e_j$ is defined as 
\begin{equation}
  c^{pe}_{ij} =
    \begin{cases}
      1.0 & \text{if $g^p_i$ and $g^e_j$ co-occurred in D}\\
      \epsilon & \text{otherwise}
    \end{cases},
\end{equation}
where $\epsilon$ is a small number.
Similarly, the adjacency matrix $\mathbf{C}^{er}$ is constructed by considering the co-occurrence relationship between goal entities and goal attributes.
{
In our work, the adjacency matrix can help infer the corresponding entities by the goal type information.}
\end{defi}

\noindent \textbf{Problem statement.}
Given a dialog context $\mathbf{X}$ of previous utterances and the predefined final recommendation goal $g_f$, 
we aim to simultaneously plan multi-level goal sequences $\mathbf{g}^p$, $\mathbf{g}^e$, and $\mathbf{g}^r$
to proactively and naturally lead the conversation to reach the final goal.
Based on the item recommendation requirement of CRS, we regard the entity level goal planning as the main task, the type level and attribute level goal planning as auxiliary tasks.
%

%

\section{Related work}
This work is related to conversational recommender system and hierarchical structure modeling.
%

\textbf{Conversational recommender system.}
The research on the conversational recommender system has two research lines.
One is from the recommender system and another is from the dialogue system.
The research works of the first one~\cite{sun2018conversational, christakopoulou2018q, zhang2018towards, yu2019visual, lei2020estimation, lei2020interactive, zhang2022multiple} aim to infer user's interest by historical interactions and the system generally consider two actions: ask questions or recommend items.
The second line from the dialogue system aims to enforce natural semantic transitions in multi-turn human-machine natural language interactions.
For example, 
\cite{xu2020conversational, liu2022graph} proposed to leverage the information of global graph structure to enhance goal embedding learning, and \cite{liu-etal-2020-towards-conversational,zhou2020topicguided} incorporated topic threads to enforce natural semantic transitions towards recommendation.
\color{black}
Additionally, \cite{li2023trea} builds a scalable, multi-layered tree to elucidate causal links between entities and leverages past conversations for more relevant recommendation responses.
Meanwhile, \cite{mao2023unitrec} uses a dual-attention Transformer encoder to capture user history context.
Other notable contributions like \cite{shin-etal-2023-pivotal} and \cite{rubin-etal-2023-entity} center their research on refining representation learning from language and entity perspectives, respectively, thus advancing the capabilities of conversational recommendation.

Our work falls under the second research line. Distinctively, we pinpoint an underexplored area in this domain: the untapped potential of recognizing dialog goals across various granularities. The main problem our paper aims to address in CRS is the limited utilization of dialog goals that can be observed simultaneously at varying levels. To bridge this gap, we introduce an approach to exploit multi-level goal sequences and their inherent hierarchical relationships to enhance conversational recommendation.
\color{black}

\textbf{Hierarchical structure modeling.}
Hierarchical structure modeling has attracted lots of research attention in  many fields including recommender system~\cite{xu2020deep,qi-etal-2021-hierec} and natural language processing~\cite{su-etal-2021-dialogue, hu-etal-2021-r2d2, wu2021hi, chen-etal-2021-hierarchy, wang-etal-2021-concept}.
To name a few,
\cite{xu2020deep, qi-etal-2021-hierec} model the user's interest hierarchy from a higher level to a lower level. 
%
%
To train the matching model in an "easy-to-difficult" scheme, \cite{su-etal-2021-dialogue} proposed a hierarchical curriculum learning framework that consists of the corpus-level curriculum and the instance-level curriculum.
\cite{hu-etal-2021-r2d2} introduced a recursive transformer to model multiple levels of granularity (e.g., words, phrases, and sentences)
and \cite{wu2021hi}
proposed a Hi-Transformer which models documents in a hierarchical way.
%

In this paper, we exploit the hierarchical structure of goal sequences from dual spaces, including representation space and optimization space.
%
%

\section{Method}
%

\subsection{Framework Overview}
\begin{figure*}[htbp]
    \centering
    \includegraphics[width=.950\textwidth]{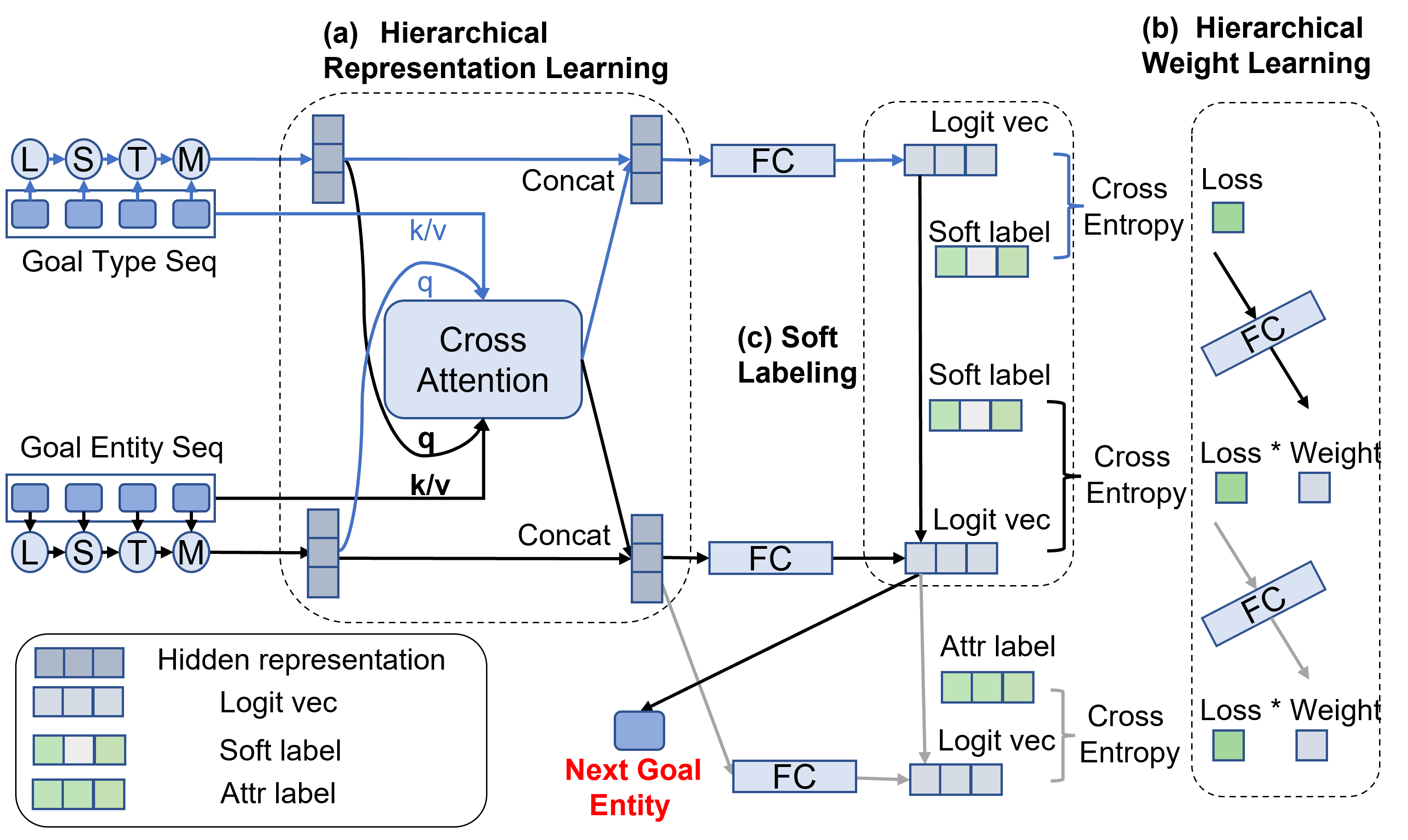}
    \caption{Overview of DHL.
    Taking the goal type sequence and the goal entity sequence as input, DHL outputs the next goal entity to lead the conversation via dual-space hierarchical learning and soft labeling guidance.
    }
    \label{fig: HGG}
\end{figure*}

Figure~\ref{fig: HGG} shows an overview of DHL that includes the following three tasks, 
(1) learning enhanced multi-level goal representations in a shared latent {representation} space, 
(2) robustly optimizing goal representations by exploiting cross-level supervision signals {in the optimization space}, and
(3) guiding the optimization direction of goal sequences toward the final recommendation goal.
{The first two tasks study the multi-level goal sequences in two spaces and the last task exploits the single-level goal sequence dependency to guide the recommendation to reach the final goal.
Specifically,
}
for the first task, we construct a two-layer goal sequence hierarchy between goal type and goal entity, and propose the \textit{hierarchical representation learning} to obtain goal representations by capturing cross-level dependencies in a mutual reinforcement way.
For the second task, we propose the \textit{hierarchical weight learning} to achieve robust goal representation optimization by reweighting multi-level goal sequences via bi-level optimization.
For the third task, we propose a \textit{soft labeling} strategy to gradually enforce the global optimization direction by attaching information of the final target to the goal sequence.

\subsection{Base Model}
We adopt LSTM~\cite{hochreiter1997long} as the basic building block for proactive goal planning, with the consideration of sequential relationship in the dialog goal entity sequence.

We derive the hidden representation of the goal entity sequence by $\mathbf{h}_i^e=\rm{LSTM}(\mathbf{x}_i,\mathbf{h}_{i-1}^e)$ where $\rm{LSTM(\cdot,\cdot)}$ is the LSTM function.
\textcolor{black}{Here $\mathbf{x}_i$ is the fixed-length initialization for the $i_{th}$ goal and $i$ ranges from $1$ to $t$.}
Then we can derive the logit of the next goal entity by
\begin{equation}\label{equ:score}
\mathbf{l}_{t+1}^e= \rm{MLP}(\mathbf{h}_t^e),
\end{equation}
where $\mathbf{l}_{t+1}^e$ is a logit vector for all goals, and $\rm{MLP}(\cdot)$ is a fully connected neural network.
We optimize the cross-entropy loss $\mathcal{L}_e(\boldsymbol{\theta})$ between the one-hot encoded next goal entity $\mathbf{g}^e_{next}$ and the goal entity probability vector $\rm{softmax(\mathbf{l}^e)}$,
where $\boldsymbol{\theta}$ are learnable model parameters.

\subsection{Hierarchical Representation Learning}
\label{subsec: cross_attention}
%

Then we introduce the hierarchical representation learning to further exploit relationships between multi-level goal sequences.
The multi-level goal sequences describe dialog topics in multiple granularities (illustrated in Definition \ref{def:goal}
{\textbf{Goal}}), which can be exploited to improve goal prediction.
For instance, if the logit score of goal type \textit{Music Rec} is high and it ever co-occurred with goal entity \textit{Cut My Heart} in historical dialogs, the likelihood of the goal entity \textit{Cut My Heart} should be relatively larger than others.
To incorporate such hierarchical knowledge, we integrate high-level information into low-level tasks based on $\mathbf{C}^{pe}$ and $\mathbf{C}^{er}$:
\begin{equation}
\label{eqn: hier}
\begin{aligned}
    \mathbf{l}^e &= \mathbf{l}^e + \rm{softmax}(\mathbf{l}^p) \cdot \mathbf{C}^{pe}, \\
    \mathbf{l}^r &= \mathbf{l}^r + \rm{softmax}(\mathbf{l}^e) \cdot \mathbf{C}^{er},
\end{aligned}
\end{equation}
where $\mathbf{l}^p\in\mathbb{R}^{N_p}$, $\mathbf{l}^e\in\mathbb{R}^{N_e}$ and $\mathbf{l}^r\in\mathbb{R}^{N_r}$ are the logit vectors of goal type, the goal entity and the goal attribute derived from Eq.~(\ref{equ:score}), and $\cdot$ is the matrix multiplication operation.
The next goal type, the next goal entity and the next goal attribute can be calculated as the one with the largest logit in $\mathbf{l}^{p}$, $\mathbf{l}^{e}$ and $\mathbf{l}^{r}$, respectively.

Based on the hierarchical information exchange design, we propose the \textit{cross attention} module to learn enhanced goal representations, which including two symmetric components: the type2entity attention component and the entity2type attention component.
Specifically, the type2entity attention component adaptively absorbs the goal type knowledge during the goal entity sequence learning, and the entity2type attention component dynamically integrates the goal entity knowledge during the goal type sequence learning. 
Due to page limit, we use the type2entity attention component for illustration, and the entity2type 
works in a similar way.

Formally, consider the sequence of goal entity embeddings $[\mathbf{x}^e_1, \mathbf{x}^e_2, \dots, \mathbf{x}^e_t]$ and the hidden state of the goal type sequence  $\mathbf{h}^p_t$ derived from the LSTM encoder.
We transform the sequence of goal entity embeddings as the sequence of key embeddings $[\mathbf{x}^k_1, \mathbf{x}^k_2, \dots, \mathbf{x}^k_t]$ and value embeddings $[\mathbf{x}^v_1, \mathbf{x}^v_2, \dots, \mathbf{x}^v_t]$ by MLPs and treat $\mathbf{h}^p_t$ as the query embedding.
The distilled knowledge representation by the type2entity component is defined by
\begin{equation}
    \mathbf{h}^{pe} = \sum_{j=1}^t \frac{f({\mathbf{h}^p_t, \mathbf{x}^k_j})}{\sum_{l=1}^t f(\mathbf{h}^p_t, \mathbf{x}^k_l)}\mathbf{x}^v_j,
\end{equation}
where $f({\mathbf{h}^p_t, \mathbf{x}^k_j})$ represents the exponential kernel $exp(\frac{\mathbf{h}^p_t(\mathbf{x}^k_j)^T}{\sqrt{d}})$ and $d$ denotes the embedding size.
Then we simply concatenate $\mathbf{h}^{pe}$ with $\mathbf{h}^e_t$ to integrate the distilled knowledge into the next goal entity prediction.

\subsection{Hierarchical Weight Learning}
Besides capturing the hierarchical relationship in the representation space, we also propose the hierarchical weight learning to leverage the goal sequence hierarchy in the optimization space.

A naive optimization scheme is to compute the goal type loss $\mathcal{L}_p(\boldsymbol{\theta})$,  the goal entity loss $\mathcal{L}_e(\boldsymbol{\theta})$ and the goal attribute loss $\mathcal{L}_r(\boldsymbol{\theta})$ as the cross-entropy loss between the one-hot encoded next goal $\mathbf{g}_{next}$ and the goal probability vector $\rm{softmax(\mathbf{l})}$.
Then the overall learning objective $\mathcal{L} = \mathcal{L}_p(\boldsymbol{\theta}) + \mathcal{L}_e(\boldsymbol{\theta}) + \mathcal{L}_r(\boldsymbol{\theta})$ can be optimized by gradient methods such as Adam~\cite{kingma2014adam}.
Yet, this optimization scheme neglects the hierarchical relationship in the optimization space.

\noindent \textbf{Hierarchical weight.} Intuitively, an accurate prediction of the high-level goal~(e.g., the goal type) can provide much useful information to guide the optimization of the low-level prediction task~(e.g., the goal entity prediction), which can be leveraged to improve model training.
Therefore, we introduce the hierarchical weight for the low-level prediction task, where a more accurate high-level prediction indicates a larger hierarchical weight for the low-level prediction task.

However, assigning a scalar weight for each goal sequence is not scalable due to the large number of goal sequences in real-world datasets. Therefore, we approximate the weight for each goal sequence via a neural network.
Inspired by \cite{han2018coteaching, chen2021generalized, chen2022unbiased}, we adopt an MLP network followed by a sigmoid function to output the hierarchical weight. 
Specifically, the MLP network takes the loss of goal type as input and outputs the hierarchical weight for the goal entity loss. 
And the same weight assignment operation can be applied for goal attributes based on goal entities.
Denote the MLP parameters as $\boldsymbol{\alpha}$, the hierarchical weighted loss function is

\begin{small}
\begin{equation}
    \label{eqn: weighted_loss}
    \!\mathcal{L}(\boldsymbol{\theta}, \boldsymbol{\alpha}) \!= \!\mathcal{L}_p(\boldsymbol{\theta})\!+\!\omega_e^{\boldsymbol{\alpha}}(\mathcal{L}_p)\mathcal{L}_e(\boldsymbol{\theta})\!+\! \omega_r^{\boldsymbol{\alpha}}(\mathcal{L}_e)\mathcal{L}_r(\boldsymbol{\theta}),
\end{equation}
\end{small}
where $\omega_e^{\boldsymbol{\alpha}}(\mathcal{L}_p)$ and $\omega_r^{\boldsymbol{\alpha}}(\mathcal{L}_e)$ denote the hierarchical weights for the goal entity loss and the goal attribute loss, respectively. 
In the following, we denote $\omega_e^{\boldsymbol{\alpha}}(\mathcal{L}_p)$ as $\omega_e^{\boldsymbol{\alpha}}$ and $\omega_r^{\boldsymbol{\alpha}}(\mathcal{L}_e)$ as $\omega_r^{\boldsymbol{\alpha}}$ for brevity.

\noindent \textbf{Bi-level optimization for weight learning.} 
Directly optimizing Eq.~(\ref{eqn: weighted_loss}) may lead to a trivial solution where all hierarchical weights reduce to zeros. In this work, we propose to formulate hierarchical weight learning as bi-level optimization~\cite{franceschi2018bilevel, chen2022gradient, yuan2023importance} where the hierarchical weight is decided by an outer level learning task:

\begin{small}
\begin{equation}
\begin{aligned}
    \min_{\boldsymbol{\alpha}} \  & \mathcal{L}_{out}\!=\! \mathcal{L}_p(\boldsymbol{\theta}^*(\boldsymbol{\alpha})) \!+\!\mathcal{L}_e(\boldsymbol{\theta}^*(\boldsymbol{\alpha}))\!+\!\mathcal{L}_r(\boldsymbol{\theta}^*(\boldsymbol{\alpha})). &\\
\!\mbox{s.t.}\ \ \!&\! \boldsymbol{\theta}^*(\boldsymbol{\alpha})\!=\! \mathop{\arg\min}_{\boldsymbol{\theta}}  \mathcal{L}_p(\boldsymbol{\theta})\!+\!\omega_e^{\boldsymbol{\alpha}}\mathcal{L}_e(\boldsymbol{\theta})\!+\!\omega_r^{\boldsymbol{\alpha}}\mathcal{L}_r(\boldsymbol{\theta}).
\end{aligned}
\end{equation}
\end{small}

\noindent In this formulation, the inner variable is the model parameters $\boldsymbol{\theta}$ and the outer variable is the MLP network parameters $\boldsymbol{\alpha}$.
We build the connection between $\boldsymbol{\theta}$ and $\boldsymbol{\alpha}$ in the inner loop via a gradient descent step, and optimize $\boldsymbol{\alpha}$ in the outer loop.
Denote $\mathcal{L}_{in} = \mathcal{L}_p(\boldsymbol{\theta}) + \omega_e^{\boldsymbol{\alpha}}L_e(\boldsymbol{\theta})+\omega_r^{\boldsymbol{\alpha}}\mathcal{L}_r(\boldsymbol{\theta})$.
For the the inner level loop, we have:
\begin{equation} \label{inner}
\begin{aligned}
    \boldsymbol{\theta}^{*}(\boldsymbol{\alpha})  \approx \boldsymbol{\theta} - \eta \frac{\partial{\mathcal{L}_{in}(\boldsymbol{\theta}, \boldsymbol{\alpha})}}{\partial{\boldsymbol{\theta}}}.
\end{aligned}
\end{equation}
In the outer level, we update $\boldsymbol{\alpha}$ by minimizing ${\mathcal{L}}_{out}(\boldsymbol{\theta}^*(\boldsymbol{\alpha}))$ via the gradient descent method with the learning rate $\eta^{'}$:
\begin{equation} \label{outer}
\begin{aligned}
    \boldsymbol{\alpha}^* \approx  \boldsymbol{\alpha} - \eta^{'} \frac{\partial{\mathcal{L}_{out}(\boldsymbol{\theta}^*(\boldsymbol{\alpha}))}}{\partial{\boldsymbol{\alpha}}}.
\end{aligned}
\end{equation}
In this way, we can leverage the weak supervision signals derived from the outer level task to update hierarchical weights stably.
In the final stage, we calculate the weighted loss term in Eq.~(\ref{eqn: weighted_loss}) by using the newly updated hierarchical weights $\omega_e^{\boldsymbol{\alpha}}$ and $\omega_r^{\boldsymbol{\alpha}}$.

\subsection{Soft Labeling}
\begin{figure}
\centering
\begin{minipage}[t]{1.0\textwidth}
  \centering
    \captionsetup{width=.95\linewidth}
    \includegraphics[width=0.8\columnwidth]{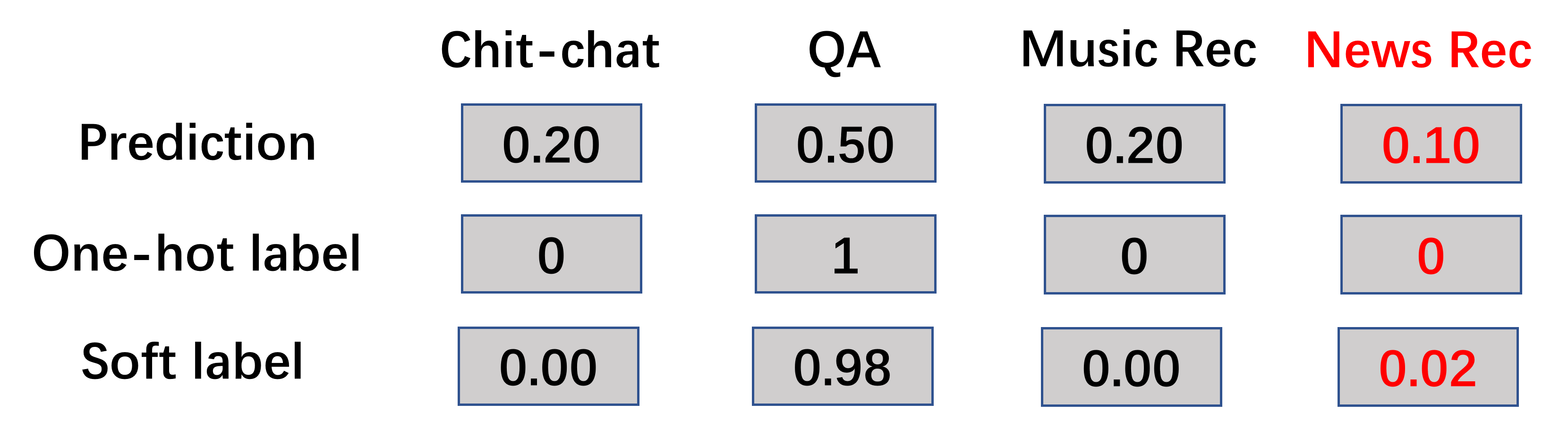} 
    \caption{
    An illustrative example of the soft labeling strategy.
    A soft parameter $0.02$ is attached to the one-hot goal label, which guides the prediction to approach the final recommendation goal \textit{News Rec}.
    }
    \label{fig: soft_label_example}
\end{minipage}%
\end{figure}
As the number of goals is finite, the next goal prediction task can be formulated as a multi-class classification task. 
Traditional classification tasks optimize the cross-entropy loss between the probability score and the one-hot label.
Different from such formulation, one unique characteristic of the goal guided conversational recommender system is to proactively and naturally lead the user to the final recommendation goal when planning the short-term goal in each step.

In this work, we achieve gradual guidance by introducing the \textit{soft labeling} strategy,
which enforces the prediction closer to the final goal as the conversation went on.
As shown in Figure~\ref{fig: soft_label_example}, we attach the one-hot goal label with a soft parameter in the final goal position.
As the goal sequence unrolls, we gradually increase the soft parameter.
The soft parameter $s_p$ is calculated as 
\begin{equation}
    \label{eqn: soft_l}
    s_p = s_0 \min~(\frac{L}{10}, 1),
\end{equation}
where $L$ is the goal sequence length and $s_0 = 0.02$ is a hyper-parameter that controls the strength of soft labeling.
From Eq.~(\ref{eqn: soft_l}), 
We can observe that $s_p$ increases as $L$ increases and it remains unchanged when $L$ is larger than $10$.
Denote the soft label of goal entity as $\mathbf{g}^e_{soft}$, the goal entity loss can be computed by $\mathcal{L}_e(\boldsymbol{\theta}) = \rm{CE}(\mathbf{g}^e_{soft}, \rm{softmax}(\mathbf{l}^e))$
where $\rm{CE(\cdot)}$ is the cross entropy loss.
The loss function for other levels of goals can be computed similarly.
By leveraging the proposed soft labeling strategy, the goal sequence transits to the final recommendation goal gradually.

\section{Experiments}

\subsection{Datasets}
We perform experiments on two datasets: DuRecDial~\cite{liu-etal-2020-towards-conversational} and TG-ReDial~\cite{zhou2020topicguided}.
%
%
%
%
%
%
%
Note that there are type level, entity level and attribute level goals in DuRecDial  while there are no explicit goal attributes in TG-ReDial.
Therefore, we only consider goal type prediction as the auxiliary task in the experiments on TG-ReDial.
See Appendix~\ref{sec:Dataset} for the details of datasets.

\subsection{Baselines}
We have chosen five baselines for comparison:

\begin{itemize}
    \item \textbf{Next} chooses the last goal entity as the prediction of the next one; 
    \item \textbf{LSTM} leverages an LSTM to model goal entity sequence dependency;
    \item \textbf{CNN}~\cite{liu-etal-2020-towards-conversational} proposes a CNN model for the goal entity sequence learning;
    \item \textbf{TG}~\cite{zhou2020topicguided} uses the dot product between the learned representation and every goal entity embedding for ranking;
    \item \textbf{MGNN}~\cite{xu2020conversational, liu2022graph} is the state-of-the-art graph neural network based model for dialog policy learning. 
\end{itemize}

\subsection{Implementation Details}
%
For all methods, we set the embedding size of goal type, goal entity and goal attribute as $256$ and the batchsize as $128$.
%
%
We use the dev dataset to tune other hyper-parameters such as the learning rate.
%
%
%
%
%
%
%
%
\textcolor{black}{To assess the performance of models in goal entity prediction, we utilize a range of metrics, namely accuracy (Acc), recall (Rec), precision (Prec), and F1 scores. These metrics provide a holistic view of each model's efficacy. Further details and clarifications on these evaluation metrics can be found in \ref{sec: evaluation}.}
%
%
Besides, we use the dialog-leading success rate metric(LS) to measure how well a model can lead the dialog to approach the final recommendation goal.
More specifically, LS is the ratio of the number of achieved final recommendation goals over the number of all goal predictions.
Please refer to ~\ref{sec:Training} for more training details.

\subsection{Overall Results}

\begin{table}
\centering
\caption{Overall Results on DuRecDial Dataset.}
\scalebox{1.0}{
\begin{tabular}{ccccccc}
\toprule
Metrics & Next &LSTM &CNN & TG & MGNN & DHL \\
\midrule
Acc(\%) & $68.75$& $78.24$& \underline{$81.85$} & $79.48$ &$67.31$ & $\textbf{83.51}$\\
Rec(\%) & \underline{$59.34$}& $46.56$& $55.84$ & $51.21$ &$25.95$ & $\textbf{62.62}$\\
Prec(\%) & \underline{$66.34$}& $54.33$& $61.82$ & $61.79$ &$26.81$ & $\textbf{72.52}$\\
F1(\%) & \underline{$61.37$}& $48.29$& $56.50$ & $54.31$ &$23.91$ & $\textbf{65.25}$\\
LS(\%) & $11.49$& $13.38$& $12.87$ & \underline{$14.22$} &$\textbf{14.46}$ & $13.52$ \\
\bottomrule
\end{tabular}
 }
\label{table: DuRecDial_Comparison}
\end{table}

\begin{table}
\centering
\caption{Overall Results on TGReDial Dataset.}
\scalebox{1.0}{
\begin{tabular}{ccccccc}
\toprule
 Metrics & Next &LSTM &CNN & TG & MGNN & DHL \\
\midrule
Acc(\%) & $33.59$&$37.98$& \underline{$38.78$} & $37.06$ &$36.98$ & $\textbf{39.99}$\\
Rec(\%) & $15.17$&\underline{$18.19$}& $17.60$ & $14.10$ &$12.56$ & $\textbf{20.78}$\\
Prec(\%) & \underline{$29.18$}&$24.26$& $15.63$ & $8.98$ &$7.76$ & $\textbf{46.32}$\\
F1(\%) & $12.76$& \underline{$17.68$}&$14.45$ & $10.23$ &$8.74$ & $\textbf{20.43}$\\
LS(\%) & $0.00$&$9.26$& \underline{$20.16$} & $19.29$ &$2.79$ & $\textbf{24.82}$\\
\bottomrule
\end{tabular}
}
\label{table: TGReDial_Comparison}
\end{table}

Table \ref{table: DuRecDial_Comparison} and Table \ref{table: TGReDial_Comparison} summarize the results for all methods where the best results are in bold and the second-best results are marked by underlines.
Firstly, we can observe that DHL outperforms all comparison methods in all metrics except for the LS metric on DuRecDial .
TG and MGNN perform slightly better in terms of the LS metric but get relatively low performance in other metrics.
We increase $s_0$ in Eq.~(\ref{eqn: soft_l}) from $0.02$ to $0.20$ in DHL and get better performance than TG and MGNN in terms of all metrics on DuRecDial ~(Accuracy: $82.65$\%; Recall: $61.56$\%; Precision: $65.45$\%; LS: $17.00$\%).
Furthermore, we notice that the Next method gets much better performance in DuRecDial  compared with the experimental results in TG-ReDial.
The reason is that the goal entity sequence in DuRecDial  contains more consecutive goal entities than that in TG-ReDial, and thus setting the last goal entity as the prediction of the next goal entity prediction performs better in DuRecDial .
%
Last but not least, CNN serves as a good baseline in both datasets.
One possible explanation is that the next goal entity prediction highly relies on the last several goal entities and CNN can model this relation well via a small sliding window.
We also analyze the computational efficiency of all comparison methods in ~\ref{sec: Computational}.

\begin{table}
\centering
\caption{Ablation Study on DuRecDial Dataset.}
\scalebox{1.0}{
\begin{tabular}{cccccccc}
\toprule
 Metrics & base & w/o att & w/o weight & w/o soft & DHL \\
\midrule
Acc(\%) & $78.24$& $82.54$ &  \underline{$82.95$}& $82.80$ &$\textbf{83.51}$\\
Rec(\%) & $46.56$& $60.82$ & \underline{$61.34$} & $60.27$ &$\textbf{62.62}$\\
Prec(\%) &$54.33$& \underline{$69.84$} & $69.20$& $67.81$ &$\textbf{72.52}$\\
F1(\%) & $48.29$&\underline{$62.98$}&  $62.90$& $61.73$ &$\textbf{65.25}$\\
LS(\%) &$13.38$& \underline{$13.46$} & $13.24$& $12.95$ &$\textbf{13.52}$\\

\bottomrule
\end{tabular}
}
\label{table: DuRecDial_Ablation}
\end{table}

\begin{table}
\centering
\caption{Ablation Study on TGReDial Dataset.}
\scalebox{1.0}{
\begin{tabular}{cccccccc}
\toprule
 Metrics & base & w/o att & w/o weight & w/o soft & DHL \\
\midrule
Acc(\%) & $ 37.98$& \underline{$39.19$} & $38.88$ &$39.09$&$\textbf{39.99}$\\
Rec(\%) & $ 18.19$& $20.00$ &  $20.17$&\underline{$20.72$}&$\textbf{20.78}$\\
Prec(\%) & $ 24.26$& $30.55$ &  $24.10$&\underline{$33.96$}&$\textbf{46.32}$\\
F1(\%) & $ 17.68$& $19.31$&  $18.80$ &\underline{$20.19$}&$\textbf{20.43}$\\
LS(\%) & $ 9.26$& $16.12$ &  $20.82$&\underline{$22.35$}&$\textbf{24.82}$\\
\bottomrule
\end{tabular}
}
\label{table: TGReDial_Ablation}
\end{table}



%








\subsection{Ablation Study}
%
We have also conducted additional experiments on both DuRecDial  and TG-ReDial with ablation consideration. 
More specifically, we remove the hierarchical representation learning, the hierarchical weight learning, and the soft labeling in DHL, and denote these variants as \textit{w/o att}, \textit{w/o weight}, \textit{w/o soft} respectively.
Besides, we denote DHL without all the three designs as \textit{base}.
The results are reported in Table \ref{table: DuRecDial_Ablation} and Table \ref{table: TGReDial_Ablation}.
%
%

Firstly, we can observe that DHL outperforms \textit{base} by a large margin~($5.27$\%  accuracy, $16.06$\% recall, $18.19$\% precision, $16.96$\% F1 and $0.14$\% LS) in DuRecDial  and~($2.01$\%  accuracy, $2.59$\% recall, $21.96$\% precision, $2.75$\% F1 and $15.56$\% LS) in TG-ReDial.
This validates the effectiveness of the proposed DHL.
Secondly, we can observe that the hierarchical representation learning and the hierarchical weight learning both play important roles in DHL. 
In the DuRecDial experiments, the precision score drops $2.68$\% without the hierarchical representation learning and drops $3.32$\% without the hierarchical weight learning.
A similar trend is observed in TG-ReDial experiments: $15.77$\% and $22.22$\% precision score drop after removing the hierarchical representation learning and the hierarchical weight learning respectively.
Last not but least, as shown in Table \ref{table: DuRecDial_Ablation} and Table \ref{table: TGReDial_Ablation}, DHL gets lower LS after removing the soft labeling strategy~(from $13.52$\% to $12.95$\% in the DuRecDial experiments and from $24.82$\% to $22.35$\% in the TG-ReDial experiments).
This can be explained as the lack of the gradual guidance to the final recommendation goal.
We further notice that the other four metrics degrade slightly~(accuracy from $83.51$\% to $82.80$\% in the DuRecDial experiments and from $39.99$\% to $39.09$\% in the TG-ReDial experiments) and the reason may be the soft labeling strategy also acts as an implicit regularization which improves the model training~\cite{szegedy2016rethinking}.

\subsection{Parameter Sensitivity}
\begin{figure}[t]
    \begin{minipage}{1.0\linewidth}
    \centering
    \hspace{-2ex}
    \subfigure[{\small Soft parameter}] {\label{fig: soft}
    \includegraphics[width=0.49\textwidth]{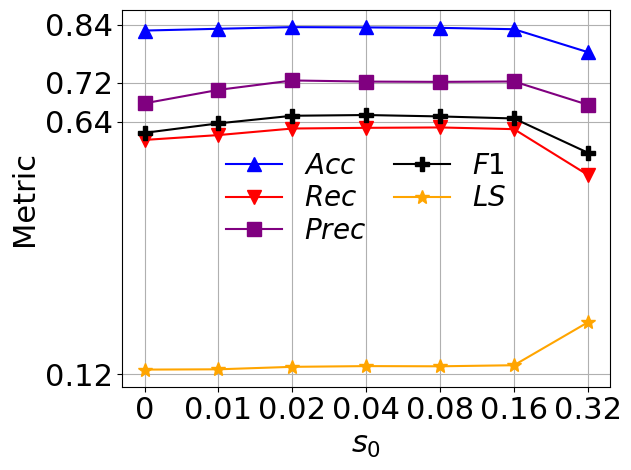}}
    \subfigure[{\small Learning rate}] {\label{fig: lr}
    \includegraphics[width=0.49\textwidth]{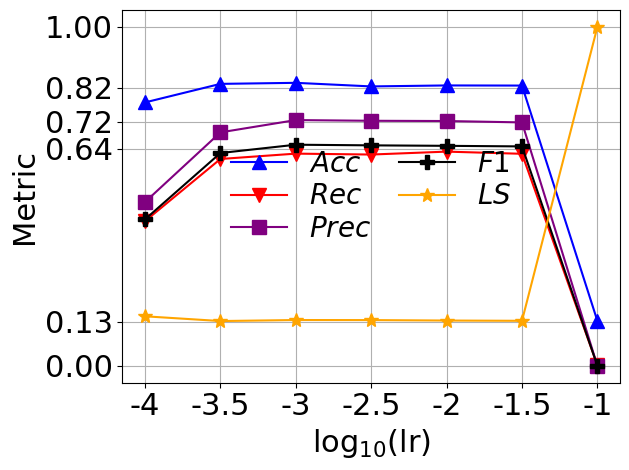}}
    \end{minipage}
    \caption{%
    Parameter sensitivity of DHL.
    %
    %
    %
    %
    }
    \label{fig: parameter}
\end{figure}
We further analyze the parameter sensitivity of DHL in DuRecDial.

\noindent \textbf{Soft parameter}.
We first examine the soft parameter and results are shown in Figure~\ref{fig: soft}.
It can be observed that as $s_0$ increases from $0$ to $0.32$, the LS score increases from $12.95$\% to $22.70$\%.
This verifies that the soft labeling strategy can lead the conversation to the final recommendation goal.
Other metrics including the accuracy, the recall, the precision and the F1 score, increase at first and then decrease as $s_0$ becomes larger than $0.02$.
The increase may be explained by that the soft labeling serves as an implicit regularization to improve the model training~\cite{szegedy2016rethinking} while the reason for the decrease is that the soft label gradually becomes a noisy label when $s_0$ becomes large.
%
We choose $s_0 = 0.02$ in our experiments.

\noindent \textbf{Learning rate}.
Besides, we examine the sensitivity of the learning rate.
As shown in Figure~\ref{fig: lr}, 
%
%
given a small learning rate of $1$e$-4$, the model gets poor performance.
As the learning rate increases from $1$e$-3.5$ to $1$e$-1.5$, the model performs well and is not sensitive to the learning rate during this period.
It can be observed that the $1$e$-3$ learning rate performs best for most metrics.
When the learning rate becomes large as $1$e$-1$, the model collapses and all goal entity predictions become the final recommendation goal.
%



\subsection{Case Study}

\begin{figure}
    \centering
    \subfigure[\small Visualization of cross attention weights.]{
    \begin{minipage}[b]{0.9\textwidth}
    \includegraphics[width=.95\textwidth]{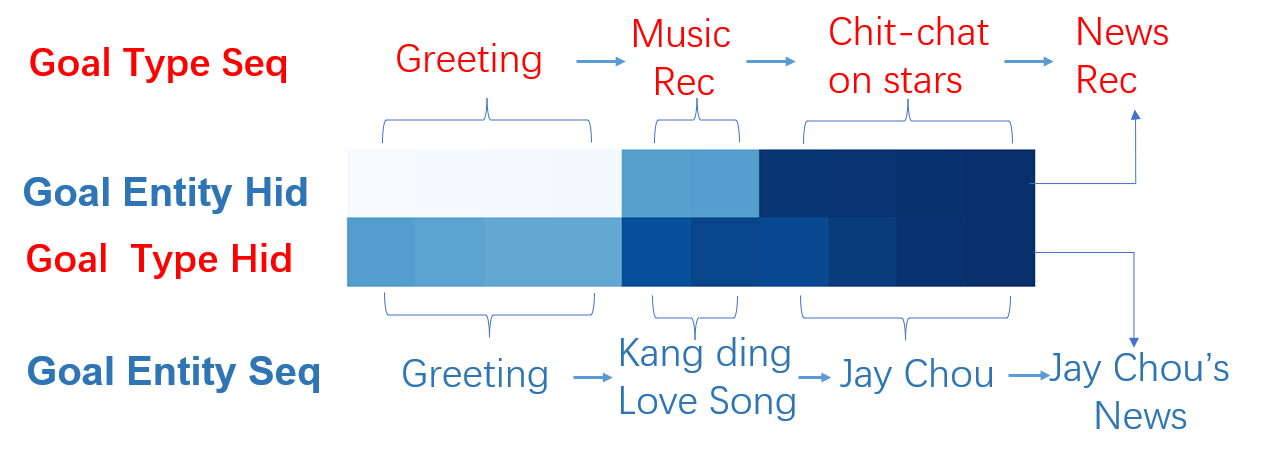}
    \label{fig: attention}
    \end{minipage}
    }
    \subfigure[\small Visualization of hierarchical weights.]{
    \begin{minipage}[b]{0.9\textwidth}
    \includegraphics[width=.95\textwidth]{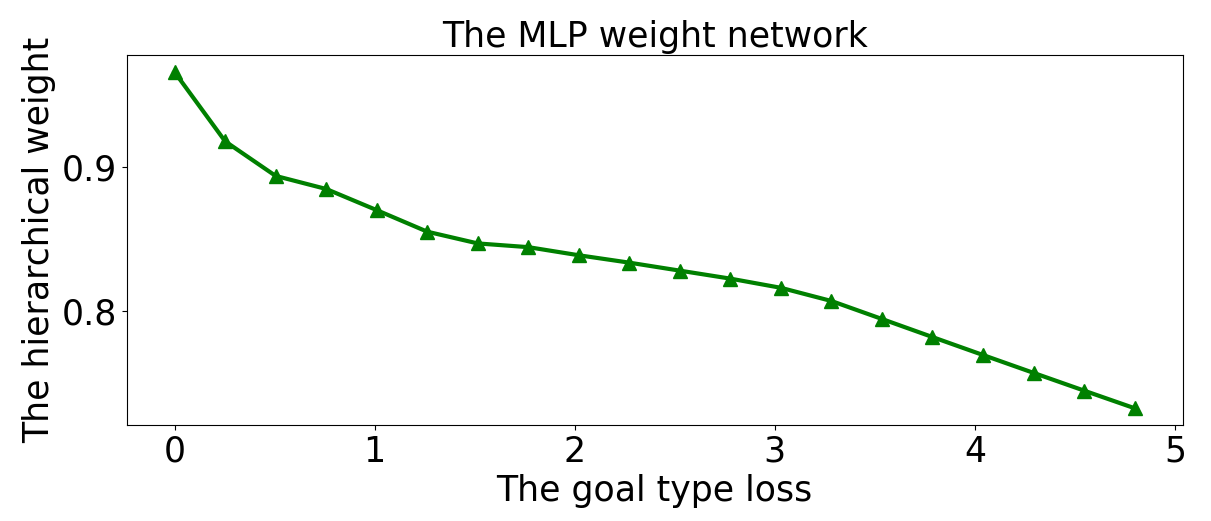}
    \end{minipage}
    \label{fig: weight}
    }
    \caption{Case study.}
\end{figure}

To qualitatively analyze the effectiveness of DHL, we visualize the cross attention weights and the hierarchical weights in DuRecDial .
%
%
%
%

\noindent \textbf{Cross attention weights}.
%
%
We extract the goal type sequence and the goal entity sequence from a dialog.
%
%
As shown in Figure~\ref{fig: attention}, we list the goal type sequence and the goal entity sequence, and visualize the cross attention weights between the hidden state and the goal embeddings.
%
%
%
%
%
We can observe the cross attention weight, between the goal type hidden state and the goal entity embeddings, increases as the goal sequence unrolls.
%
This makes sense, since the last goal entities have more influence on the next goal entity prediction.
To be specific, highlighted by the goal type hidden state,  the goal entity \textit{Jay Chou} contributes a lot to the next goal entity prediction as \textit{Jay Chou's news}.
Similarly, stressed by the goal entity hidden state, the goal type \textit{Chit-chat on Stars} contributes much to the next goal type prediction as \textit{News Rec}.
The above observations further validate the effectiveness of the hierarchical representation learning module.

\noindent \textbf{Hierarchical weights}.
%
As shown in Figure \ref{fig: weight}, 
%
we visualize the MLP weight network and the hierarchical weight for the goal entity task decreases as the goal type loss increases.
This makes sense: given a small goal type loss, the goal type prediction is accurate and thus provides much useful information for the goal entity prediction.
Therefore the goal entity prediction deserves a large task weight for optimization.
Furthermore, in the example of Figure~\ref{fig: attention}, we find the goal type loss is $0.012$ and this small loss corresponds to a large task weight $0.96$ for the goal entity prediction in Figure~\ref{fig: weight}.
Following Eq.~(\ref{eqn: hier}), we compute the goal type prediction's contribution score to the goal entity prediction as 
\begin{equation}
\begin{aligned}
      \mathbf{l}_{p2e} = \rm{softmax}(\mathbf{l}_p)\cdot \mathbf{C}^{pe}.
\end{aligned}
\end{equation}
%
The ground-truth goal entity position of $\rm{softmax}(\mathbf{l}_{p2e})$ is $1.17$\%, much larger than the average score $\frac{1}{1385} = 0.07$\%~($1385$ is the class number of the goal entity in DuRecDial).
The above results further validate the effectiveness of hierarchical weight learning.
%

\section{Conclusion}
In this paper, we propose DHL to enhance the proactive goal planning in CRS by exploiting the hierarchical structure of multi-level goal sequences.
Specifically, we propose the hierarchical representation learning in the representation space and the hierarchical weight learning in the optimization space to model hierarchical goal sequences.
%
%
%
%
%
%
In addition, we develop a novel soft labeling strategy, which can gradually guide the conversation to the final recommendation goal.
%
%
We conduct extensive experiments on DuRecDial and TG-ReDial and the results demonstrate the effectiveness of DHL.

\bibliography{anthology,custom}
\bibliographystyle{acl_natbib}


\appendix
\section{Appendix}
\label{sec:appendix}

\subsection{Dataset}

\label{sec:Dataset}


%
DuRecDial contains $10.2$k dialogues, $15.5$k utterances and $1362$ seekers and TG-ReDial contains $10.0$k dialogues, $129.4$k utterances and $1482$ seekers.
The number of goal type, goal entity, and goal attribute in DuRecDial are $22$, $1355$ and $20637$ respectively.
Following the original paper, we randomly sampling $65$\%/$10$\%/$25$\% data in DuRecDial at the level of seekers to form the train/dev/test datasets.
As for TG-ReDial, we extract the topic as the goal entity, and reduce the number of goal entity by clustering and manual processing.
The number of goal type and goal entity in TG-ReDial are $6$ and $100$ respectively.
\textcolor{black}{We treat the action in TG-ReDial as the goal type and follow the train/dev/test splitting in the original paper.
The average goal sequence length is $5.6$ in the training set.}

\color{black}
\subsection{Evaluation Metrics}
\label{sec: evaluation}

The task of goal entity prediction is fundamentally a multi-class classification problem. To evaluate the performance of our proposed methods on this task, we employ a variety of standard metrics that encompass different aspects of classification quality, namely accuracy, recall, precision, and F1 scores. Herein, we present the definitions, equations, and abbreviations for these metrics for clarity:

\begin{itemize}
    \item \textbf{Accuracy (Acc)}: Accuracy provides an overall measure of the classifier's correctness across all classes. It represents the proportion of total predictions that were correct. 
    \[ \text{Accuracy (Acc)} = \frac{TP + TN}{TP + TN + FP + FN} \]
    For multi-class problems, accuracy doesn't require averaging over classes, as it captures the overall correctness.

    \item \textbf{Recall (Rec)}: Also known as sensitivity or the true positive rate, recall, when viewed from a one-class perspective, measures the proportion of actual positive instances of that class that were correctly predicted.
    \[ \text{Recall (Rec)} = \frac{TP}{TP + FN} \]

    \item \textbf{Precision (Prec)}: Viewed from a one-class perspective, precision conveys the proportion of predicted positive instances for that class that are truly positive.
    \[ \text{Precision (Prec)} = \frac{TP}{TP + FP} \]
    
    \item \textbf{F1 Score (F1)}: The F1 score is the harmonic mean of precision and recall, providing a balance between them, especially when their values are low. Like precision and recall, the F1 score can be viewed from a one-class perspective.
    \[ \text{F1} = 2 \times \frac{\text{Precision} \times \text{Recall}}{\text{Precision} + \text{Recall}} \]
\end{itemize}

In our results, we report the macro-average of these metrics. The macro-average metric computes each metric (except accuracy) independently for each class and then takes the average.
\color{black}

\subsection{Training Details}
\label{sec:Training}
The hidden size of LSTM is set to $256$.
We adopt the cosine learning rate decay schedule for a total of 30 epochs for all comparison methods. 
We use the dev set to tune learning rate from [$1$e$-4$, $1$e$-3.5$, $1$e$-3$,  $1$e$-2.5$,  $1$e$-2$,  $1$e$-1.5$,  $1$e$-1$].
The Adam optimizer~\cite{kingma2014adam} is used in training the model parameters.
For the hierarchical weight learning, we set the embedding size of the MLP as $100$ and optimize the MLP via the Adam optimizer with a $1$e$-5$ learning rate.
We tune the soft parameter from [$0$, $0.01$, $0.02$, $0.04$, $0.08$, $0.16$, $0.32$] via the performance of the dev set.
All experiments are performed on a single Tesla P40.

\color{black}
\subsection{Computational Efficiency}
\label{sec: Computational}
As for the computational efficiency, on a single Tesla P40, MHFL takes $28.87$s to finish the inference on the whole DuRecDial test set~(LSTM: $27.85$s; CNN: $24.86$s; TG: $24.70$s; MGCG: $29.06$s).
Note that the inference speed of all methods are similar and MHFL is a little slower due to the introduction of the cross attention module.
\color{black}


\end{document}